\documentclass[12pt,a4paper]{article}

\usepackage{graphicx}
\usepackage{amsmath} 
\usepackage{natbib}
\usepackage{color}
\usepackage{enumitem}
\usepackage[hidelinks]{hyperref}
\usepackage[title]{appendix}

\usepackage[affil-sl]{authblk}

\begin{document}

\title{A constrained regression model for an ordinal response with ordinal predictors}

\author[1,2]{Javier Espinosa}
\author[2]{Christian Hennig}
\affil[1]{Department of Economics\\

USACH - Universidad de Santiago de Chile, Chile.}
\affil[2]{Department of Statistical Sciences\\

University College London, UK.}

\date{}

\maketitle

\begin{abstract}
A regression model is proposed for the analysis of an ordinal response variable depending on a set of multiple covariates containing ordinal and potentially other variables. The proportional odds model (\cite{mccullagh1980regression}) is used for the ordinal response, and constrained maximum likelihood estimation is used to account for the ordinality of covariates.

Ordinal predictors are coded by dummy variables.
The parameters associated to the categories of the ordinal predictor(s) are constrained, enforcing them to be monotonic (isotonic or antitonic). A decision rule is introduced for classifying the ordinal predictors' monotonicity directions, also providing information whether observations are compatible with both or no monotonicity direction. In addition, a monotonicity test for the parameters of any ordinal predictor is proposed. The monotonicity constrained model is proposed together with three estimation methods and compared to the unconstrained one based on simulations.

The model is applied to real data explaining a 10-Points Likert scale quality of life self-assessment variable from ordinal and other predictors.

\bigskip
{\bf Keywords:} Monotonic regression, Monotonicity direction, Monotonicity test, Constrained Maximum Likelihood Estimation.

{\bf MSC2010:} 62H12, 62J05, 62-07.
\end{abstract}

\section{Introduction}
\label{intro}
In many situations where regression models are suitable, the relationship between ordinal responses and ordinal predictors is of interest. However, statistical modelling for this type of relationship has called little attention. Even literature for ordinal predictors with any other type of scale of the response variable is scarce (see, for example, \cite{tutz2014rating}, and \cite{rufibach2010active}). 

In order to account for an ordinal response variable, proportional odds cumulative logit models (\cite{mccullagh1980regression}) are used here in presence of multiple predictors allowing for different measurement scales. We pay special attention to the treatment of ordinal scale predictors. Their parameter estimates are restricted to be monotonic through constrained maximum likelihood estimation (CMLE). To begin with, consider for simplicity one ordinal response variable $y$  with $k$ categories and one ordinal predictor $x$ with $q$ categories. The corresponding model for this setup is
\begin{align}
\text{logit}[P(y_{i}\leq j|x_i)] = \alpha_j + \sum_{p=2}^q \beta_p x_{i,p}, \label{eq:Intro_Model1}
\end{align}
$j = 1,\ldots, k-1$. $\alpha_j$ and $\beta_p$ for $p=2,\ldots q$ are real parameters.
The observations are $(x_i,y_i),\ i=1,\ldots,n$. The 
$x_{i,p}$ are dummy variables defined as $x_{i,p}=1$ if $x_i$ falls in the $p$th category of the ordinal predictor and 0 otherwise, with $p=1,\ldots, q$. Category number 1 is treated as the baseline category with $\beta_1=0$; therefore the dummy variable $x_{i,1}=1-\sum_{p=2}^q x_{i,p}$ is omitted and the sum in model \eqref{eq:Intro_Model1} starts at $p=2$. Monotonicity on $\{\beta_p\}$ is obtained by using CMLE. The general model is defined in Section \ref{sec:Model}, which allows for multiple ordinal predictors and other covariates of different measurement scales.

The monotonic effects approach to the ordinal predictors treatment is conceived here as an intermediate point between two general and common approaches within the context of regression analysis on observed variables. One of these common approaches corresponds to an unconstrained version of \eqref{eq:Intro_Model1}, treating the ordinal predictor as if it were nominal. This ignores the ordinal information.
The other common approach treats an ordinal predictor as if it were of interval scale, replacing it by a single transformed variable after applying some scoring method, $f$. More formally,
\begin{align}
\text{logit}[P(y_{i}\leq j|x_i)] = \alpha_j + \beta \tilde{x}_{i},
\end{align}
with $\tilde{x}=f(x)$. This treats $f(x)$ as interval scaled. Numerous data-based methods for scaling of ordinal variables have been proposed in the literature, on top of using plain equidistant Likert scaling (see, e.g., \cite{bross1958use}, \cite{harter1961expected}, \cite{tukey1962future}, \cite{hensler1979estimating}, \cite{brockett1981note}, and \cite{casacci2015methods}), but ultimately in most situations the data do not carry conclusive information about the appropriateness of any scaling $f$.

The intermediate approach proposed here is defined to achieve a set of linear estimates described by multiple magnitudes, as in the nominal scale approach, but allowing one direction only, as in the interval scale approach. The latter is attained by restricting the effects of the model \eqref{eq:Intro_Model1} to be monotonic in either direction.
The monotonicity assumption should not necessarily be taken for granted in regression with ordinal predictor and response. But it has a special status, similarly to linearity between interval-scaled variables. According to \cite{stevens1946measurement} the interval scale is defined by the equality in meaning of differences between values regardless of the location of these differences on the measurement range. A linear relationship between interval-scaled variables means that the impact of a change in the predictor on the response is proportional to the meaning of the change of measurement at all locations of the measurement scale. For the ordinal measurement scale, only the order of measured values is meaningful. In this situation, analogously, if the change of an ordinal predictor has impact on an ordinal response directly in line with the meaning of the change of measurements, the connection would be monotonic. 

Some other regression models for ordinal predictors are also based on the monotonic effects assumption. However, models for ordinal responses have not been explicitly discussed in this context. \cite{tutz2014rating} used penalisation methods for modelling rating scales as predictors, and an active set algorithm was proposed by \cite{rufibach2010active} to incorporate ordinal predictors in some regression models considering the response variable to be continuous, binary, or represent censored survival times, and assuming isotonic effects of the ordinal predictors' categories. Another related method is isotonic regression, mostly applied to continuous data (see, for example, \cite{barlow1972isotonic}, \cite{dykstra1982algorithm}, and \cite{stout2015isotonic}). In a broader context, there are some other types of statistical models that deal with ordinal data, such as those in item response theory (IRT) (e.g., \cite{tutz1990sequential}, \cite{bacci2014class}), latent class models (e.g., \cite{moustaki2000latent}, \cite{moustaki2003general}, \cite{vasdekis2012composite}), nonlinear principal components analysis (NLPCA) (e.g., \cite{de2009gifi}, \cite{linting2012nonlinear} and \cite{mori2016nonlinear}), and nonlinear canonical correlation analysis (NLCCA) (e.g., \cite{Mardia1979MultiAnalysis} and \cite{de2009gifi}). However, their settings are somewhat different compared to the one corresponding to modelling an ordinal response with ordinal predictors (and others) in classical linear regression. For instance, unlike IRT models and latent class models, classical regression models do not assume latent variables; and in contrast to NLPCA and NLCCA, classical regression models are not used as a dimensionality reduction technique and need a single dependent variable, respectively.

The monotonicity constrained regression model discussed here can be used for several purposes. When the unconstrained parameter estimates associated to the ordinal predictor are monotonic, then clearly there is no need of a constrained model. However, when these unconstrained estimates are non-monotonic, then there are some reasons why the constrained model could be useful. It is often of interest to compare unconstrained and constrained fits in order to decide whether there is evidence for non-monotonic relationship. 
In case that the unconstrained version does not provide a clearly better fit, the monotonic fit may be superior regarding interpretability, and may also lead to a smaller mean square error, as will be shown by simulations and a real data application.

In Section \ref{sec:Model}, the proposed model is developed in detail to obtain both constrained parameter estimates for multiple ordinal predictors and unconstrained estimates for other types of covariates. As the monotonic estimates can be either increasing (isotonic) or decreasing (antitonic), it is necessary to specify this relation while defining the constraints. Also, investigating possible directions of monotonicity for all ordinal predictors is of interest in its own right. Therefore, a monotonicity direction classification (MDC) procedure is introduced in Section \ref{sec:DecisionRule} that determines the best possible combination of isotonic and/or antitonic associations as a way of assisting the estimation method of the constrained model introduced in Section \ref{sec:Model}. In Section \ref{sec:MonoTest} a monotonicity test is proposed as a complementary tool to assess the validity of the monotonicity assumption of each ordinal predictor. Both the MDC procedure and the monotonicity test provide statistical evidence on the validity of the monotonicity assumption. This can be incorporated in the estimation procedure; Section \ref{sec:Constraints} presents two approaches, one based on the monotonicity test and another one based on the (less conservative) MDC procedure. On the other hand, the same procedures may also detect that the data are consistent with zero influence of a variable, in which case the variable may be dropped, this is treated in Section \ref{sec:MDCSelection}. Simulations are presented in Section \ref{sec:Simulations} comparing the mean square error decomposition between the constrained and unconstrained approaches. Finally, the proposed model is applied to real data from the Chilean National Socio-Economic Characterisation in Section \ref{sec:RealData}. A quality of life self-assessment variable using a 10-Points Likert scale is analysed considering ordinal and other predictors.

\section{Proportional odds with monotonicity constraints} \label{sec:Model}
For an ordinal response variable $y$ with $k$ categories, let $y_i$ be the response category for subject $i$. The model of proportional odds is 
\begin{align}
\text{logit}[P(y_{i}\leq j | \mathbf{x}_i)]=\alpha_j+\boldsymbol{\beta}'\mathbf{x}_i,
\end{align}
$j=1,\ldots, k-1,\ i=1,\ldots,n.$ A part of the elements of $\boldsymbol{\beta}$ corresponds to those effects associated to ordinal predictors categories in $\mathbf{x}$, for which their parameter estimates are constrained to account for monotonicity as explained later.

When this model has one or more of both ordinal and non-ordinal predictors, it can be represented as
\begin{equation}
\text{logit}[P(y_{i}\leq j | \mathbf{x}_i)]=\alpha_j+\sum_{s=1}^{t}\sum_{p_s=2}^{q_s}\beta_{s,p_s}x_{i,s,p_s}+\sum_{u=1}^{v}\beta_{u}x_{i,u},\label{eq:Model_eq}
\end{equation}
where $\mathbf{x}_i$ is a vector with $v-t+\sum_{s=1}^{t}q_s$ elements representing a set of $t$ ordinal predictors (OP) and their $\sum_{s=1}^t q_s$ categories together with $v$ non-ordinal predictors for the $i$th observation. Each ordinal predictor is denoted by the subindex $s$, with $s=1,\ldots,t$, and contributes $q_s-1$ dummy variables to the model representing its ordinal categories $\{1,\ldots,q_s\}$ assuming the first one as the baseline category, thus $\beta_{s,1}=0$. Each dummy variable is defined as $x_{i,s,p_s}=1$ if the $i$th observation falls in the category $p_s$ of the ordinal predictor $s$ and 0 otherwise, with $p_s=1,\ldots,q_s$. Therefore, $\mathbf{x}_i'=(x_{i,1,2},\ldots,x_{i,1,q_1},x_{i,2,2},\ldots,x_{i,2,q_2},\ldots,x_{i,t,2},\ldots,$\\$x_{i,t,q_t},x_{i,u},\ldots,x_{i,v})$, where those variables with three indexes correspond to the observation of an ordinal predictor category and those with two are observations of other types of covariates.

\subsection{Likelihood model fitting} \label{sec:lkelihood_lagrange}

Define $\pi_j(\mathbf{x}_i)=P(y_{i}=j|\mathbf{x}_i)$, the probability of the response of subject $i$ to fall in category $j$, and let $y_{i1},\ldots, y_{ik}$ be the binary indicators of the response for subject $i$, where $y_{ij}=1$ if its response falls in category $j$ and $0$ otherwise. Therefore, for independent observations, the likelihood function is based on the product of the multinomial mass functions for the $n$ subjects:
\begin{align}
&L(\{\alpha_j\},\boldsymbol{\beta})=\prod_{i=1}^n\Bigg{\{}\prod_{j=1}^k\pi_j(\mathbf{x}_i)^{y_{ij}}\Bigg{\}} \nonumber \\
&=\prod_{i=1}^n\Bigg{\{}\prod_{j=1}^k P(y_{i}=j|\mathbf{x}_i)^{y_{ij}}\Bigg{\}} \nonumber \\
&=\prod_{i=1}^n\Bigg{\{}\prod_{j=1}^k [P(y_{i}\leq j|\mathbf{x}_i)-P(y_{i}\leq j-1|\mathbf{x}_i)]^{y_{ij}}\Bigg{\}} \nonumber \\
&=\prod_{i=1}^n\Bigg{\{}\prod_{j=1}^k \Bigg{[}\frac{e^{\alpha_j+\sum_{s=1}^{t}\sum_{p_s=2}^{q_s}\beta_{s,p_s}x_{i,s,p_s}+\sum_{u=1}^{v}\beta_{u}x_{i,u}}}{1+e^{\alpha_j+\sum_{s=1}^{t}\sum_{p_s=2}^{q_s}\beta_{s,p_s}x_{i,s,p_s}+\sum_{u=1}^{v}\beta_{u}x_{i,u}}} \nonumber \\
&\quad -\frac{e^{\alpha_{j-1}+\sum_{s=1}^{t}\sum_{p_s=2}^{q_s}\beta_{s,p_s}x_{i,s,p_s}+\sum_{u=1}^{v}\beta_{u}x_{i,u}}}{1+e^{\alpha_{j-1}+\sum_{s=1}^{t}\sum_{p_s=2}^{q_s}\beta_{s,p_s}x_{i,s,p_s}+\sum_{u=1}^{v}\beta_{u}x_{i,u}}}\Bigg{]}^{y_{ij}}\Bigg{\}}.
\end{align}

Hence, 
\begin{align}
&\pi_j(\mathbf{x}_i)=\frac{e^{\alpha_j+\sum_{s=1}^{t}\sum_{p_s=2}^{q_s}\beta_{s,p_s}x_{i,s,p_s}+\sum_{u=1}^{v}\beta_{u}x_{i,u}}}{1+e^{\alpha_j+\sum_{s=1}^{t}\sum_{p_s=2}^{q_s}\beta_{s,p_s}x_{i,s,p_s}+\sum_{u=1}^{v}\beta_{u}x_{i,u}}} \nonumber \\
&\quad \quad \quad -\frac{e^{\alpha_{j-1}+\sum_{s=1}^{t}\sum_{p_s=2}^{q_s}\beta_{s,p_s}x_{i,s,p_s}+\sum_{u=1}^{v}\beta_{u}x_{i,u}}}{1+e^{\alpha_{j-1}+\sum_{s=1}^{t}\sum_{p_s=2}^{q_s}\beta_{s,p_s}x_{i,s,p_s}+\sum_{u=1}^{v}\beta_{u}x_{i,u}}},
\end{align}
and the log-likelihood function for the model is
\begin{align}
&\ell(\{\alpha_j\},\boldsymbol{\beta})=\sum_{i=1}^n\sum_{j=1}^k{y_{ij}}\log\pi_j(\mathbf{x}_i). \label{eq:likelihood_Model}
\end{align}
As we are interested in a constrained version of this model with the aim of getting monotonic increasing/decreasing effects, it is necessary to define the set of constraints to be applied on the $t$ sets of $q_s$ coefficients. The isotonic constraints are
\begin{align}
0\leq\beta_{s,2}\leq\cdots\leq\beta_{s,q_s}, \quad \forall s \in \mathcal{I},
\end{align}
where $\mathcal{I}\subseteq \mathcal{S}$, with $\mathcal{S}=\{1,2,\ldots,t\}$, and $\beta_{s,1}=0$. The antitonic constraints are
\begin{align}
0\geq\beta_{s,2}\geq\cdots\geq\beta_{s,q_s}, \quad \forall s \in \mathcal{A},
\end{align}
where $\mathcal{A}\subseteq \mathcal{S}$, and $\beta_{s,1}=0$. An estimation method based on a monotonicity direction classification (MDC) procedure will be discussed in Section \ref{sec:DecisionRule}, allocating the ordinal predictors in either of these two subsets, achieving $\mathcal{I}\cup \mathcal{A}= \mathcal{S}$.

These constraints can be expressed in matrix form as $\mathbf{C}\boldsymbol{\beta}_{(ord)}\geq \mathbf{0}$. The vector $\boldsymbol{\beta}_{(ord)}$ is part of the vector $\boldsymbol{\beta}$. The latter contains all the parameters associated to the $t$ ordinal predictors and their $q_s-1$ categories together with the $v$ non-ordinal predictors, $\boldsymbol{\beta}'=\left(\boldsymbol{\beta}_{(ord)}',\boldsymbol{\beta}_{(non-ord)}'\right)$, with $\boldsymbol{\beta}_{(ord)}'=(\boldsymbol{\beta}_1',\ldots,\boldsymbol{\beta}_t')$ with $s=1,\ldots,t,$ and $\boldsymbol{\beta}_{(non-ord)}'=(\beta_1,\ldots,\beta_v)$ with $u=1,\ldots,v$, where each vector $\boldsymbol{\beta}_s'=(\beta_{s,2},\ldots,\beta_{s,q_s})$ with $p_s=2,\ldots,q_s$. The matrix $\mathbf{C}$ is a square block diagonal matrix of $\sum_{s=1}^{t}(q_s-1)$ dimensions composed of $t$ square submatrices $\mathbf{C}_s$ in its diagonal structure and zeros in its off-diagonal blocks as follows,
$$
\mathbf{C}=\left[\begin{array}{ccccc}
\mathbf{C}_{1}&\mathbf{0}&\cdots &\mathbf{0}\\
\mathbf{0}&\mathbf{C}_{2}&\mathbf{0} &\mathbf{0}\\
\mathbf{0}& \cdots & \ddots  &\mathbf{0}\\
\mathbf{0}& \cdots & \cdots &\mathbf{C}_{t}\\
\end{array}\right], \text{ with } s=1,\ldots,t, \label{BlockDiagMatrix} $$ 
where
$$ \mathbf{C}_s=\left[\begin{array}{cccc}
1&0& \cdots &0\\
-1&1& 0 &0\\
0 &\ddots &\ddots & 0\\
0&\cdots &-1&1\\
\end{array}\right] \quad \forall s \in \mathcal{I},$$
$$ \mathbf{C}_s=\left[\begin{array}{cccc}
-1&0&\cdots &0\\
1&-1& 0 &0\\
0 &\ddots &\ddots & 0 \\
0&\cdots &1&-1\\
\end{array}\right] \quad \forall s \in \mathcal{A}, \label{BlockDiagSubMatrices} $$ 
and each square submatrix $\mathbf{C}_s$ has $q_s-1$ dimensions.

Then, the maximisation problem is
\begin{align}
\text{maximise }&\ell(\{\alpha_j\},\boldsymbol{\beta}) \nonumber \\
\text{subject to }&\mathbf{C}\boldsymbol{\beta}_{(ord)}\geq \mathbf{0}, \label{eq:MaximizationProblem}
\end{align} 
where $\mathbf{0}$ is a vector of $\sum_{s=1}^{t}(q_s-1)$ elements. Now, \eqref{eq:MaximizationProblem} can be expressed as the Lagrangian
\begin{align}
\mathcal{L}(\{\alpha_j\},\boldsymbol{\beta},\boldsymbol{\lambda})&=\ell(\{\alpha_j\},\boldsymbol{\beta})-\boldsymbol{\lambda}'\mathbf{C}\boldsymbol{\beta}_{(ord)}, \label{eq:lagrangian_Model}
\end{align}
where $\boldsymbol{\lambda}$ is the vector of $\sum_{s=1}^t(q_s-1)$ Lagrange multipliers denoted by $\lambda_{s,p_s}$.

The set of equations to be solved is obtained by differentiating $\mathcal{L}(\{\alpha_j\},\boldsymbol{\beta},\boldsymbol{\lambda})$ with respect to its parameters and equating the derivatives to zero. In order to solve this in \texttt{R}, the package \texttt{maxLik} (\cite{maxLik2011}) offers the \texttt{maxLik} function which refers to \texttt{constrOptim2}. This function uses an adaptive barrier algorithm to find the optimal solution of a function subject to linear inequality constraints such as in \eqref{eq:MaximizationProblem} (\cite{lange2010numerical}).

\section{Monotonicity direction classification} \label{sec:DecisionRule}
Under the monotonicity assumption for all OPs, an important decision to be made is whether each ordinal predictor's set of effects (also referred to as pattern), is either isotonic, namely $s\in \mathcal{I}$, or antitonic, $s\in \mathcal{A}$. Also outside the context of parameter estimation, it may be of interest whether a predictor is connected to the response in an isotonic or antitonic way, or potentially whether monotonicity may not hold or whether both directions are compatible with the data.

One possible way to deal with this decision is to just maximise the likelihood, i.e., to fit $2^{t}$ models, one for each possible combination of monotonicity directions for the $t$ ordinal predictors, and then choose the one with the highest likelihood. However, as the number of ordinal predictors $t$ increases, the number of possible combinations of monotonicity directions becomes greater, which could lead to a considerable number of high dimensional models to be fitted.

Another possible estimation method uses a monotonicity direction classifier to find the monotonicity direction for each ordinal predictor and then fits only one model. This will be based on confidence intervals (CIs) for the parameters and on checking which monotonicity direction is compatible with these. This may miss the best model, but in some situations it may be desirable to take into account fewer than $2^t$ but more than a single model.

The two approaches are put together in a three steps monotonicity direction classification (MDC) procedure exploiting their best features. Each of the first two steps uses a decision rule with different confidence levels for the CIs, and the last step applies the multiple models fitting process described above over those patterns with no single monotonicity direction established in the previous steps. Before describing its steps, consider some remarks and definitions.

The parameters' CIs from an unconstrained model are the main input for the decision rule proposed here. It is possible to compute the CI defined in equation \eqref{eq:CI} for the parameters of an unconstrained version of the model \eqref{eq:Model_eq} (\cite{agresti2010analysis}). Denote $SE_{\hat{\beta}}$ as the standard error of the parameter estimate $\hat{\beta}$, then an approximate confidence interval for $\beta$ with a $100(1-\tilde{\alpha})\%$ confidence level is
\begin{align}
\hat{\beta}\pm z_{\tilde{\alpha}/2}(SE_{\hat{\beta}}), \label{eq:CI}
\end{align}
where $z_{\tilde{\alpha}/2}$ denotes the standard normal percentile with probability $\tilde{\alpha}/2$. The values for $\hat{\beta}$ and $SE_{\hat{\beta}}$ are obtained by fitting the proportional odds model (\cite{mccullagh1980regression}) over the unconstrained model \eqref{eq:Model_eq}. The \texttt{R} function \texttt{vglm} of the package \texttt{VGAM} was used here (\cite{yee2015vgam}).

The first two steps of the MDC procedure provide four possible outcomes for each pattern of unconstrained parameter estimates associated to an ordinal predictor's categories: `isotonic', `antitonic', `both', and `none'. The first two correspond to a classification of monotonicity direction whereas the remaining two to the case where a single direction is not found because either both directions of monotonicity are possible or the parameter estimates' pattern is not compatible with monotonicity, respectively. The idea is that the intersections of all CIs for the parameters of a single ordinal predictor together will either allow for isotonic but not antitonic parameters, or for antitonic but not isotonic parameters, or for both, or for neither. 
Formally, the MDC of the parameter estimates' pattern is defined as
\begin{align}
d_{s,\tilde{c}} = 
     \begin{cases}
       \text{isotonic}  &\quad\text{if }\mathcal{D}_{s,\tilde{c}}= \{0,1\}  \text{ or }\mathcal{D}_{s,\tilde{c}}= \{1\}\\
       \text{antitonic}  &\quad\text{if }\mathcal{D}_{s,\tilde{c}}=\{-1,0\}  \text{ or }\mathcal{D}_{s,\tilde{c}}= \{-1\}\\
       \text{both}  &\quad\text{if }\mathcal{D}_{s,\tilde{c}}=\{0\}\\
       \text{none}  &\quad\text{if }\mathcal{D}_{s,\tilde{c}}\supseteq\{-1,1\}, \label{eq:_decisionRule1_2_V2}
     \end{cases}
\end{align}
where $\mathcal{D}_{s,\tilde{c}}=\{d_{s,p_s,p_s',\tilde{c}}\}$ is defined as the set of distinct values resulting from \eqref{eq:_decisionRule1_1} for the ordinal predictor $s$ considering confidence intervals with a $100\tilde{c}\%$ confidence level, and 
\begin{align}
d_{s,p_s,p_s',\tilde{c}} = 
     \begin{cases}
       1  &\quad\text{if }\tilde{L}_{s,p_s,\tilde{c}}>=\tilde{U}_{s,p_s',\tilde{c}}\\
      -1  &\quad\text{if }\tilde{U}_{s,p_s,\tilde{c}}<=\tilde{L}_{s,p_s',\tilde{c}}\\
       0  &\quad\text{otherwise,} \\ 
     \end{cases} 
\label{eq:_decisionRule1_1}
\end{align}
$\forall p_s'<p_s$ and $p_s \in \{2,3,\ldots,q_s\}$, where $\tilde{U}_{s,p_s,\tilde{c}}$ is the confidence interval's upper bound of the parameter $\beta_{s,p_s}$ associated to the category $p_s$ of the ordinal predictor $s$ given a $100\tilde{c}\%$ confidence level, and $\tilde{L}_{s,p_s,\tilde{c}}$ is its corresponding lower bound. Note that, by definition, the first category of all ordinal predictors is set to zero, so $\tilde{L}_{s,1,\tilde{c}}=\tilde{U}_{s,1,\tilde{c}}=0$, $\forall s$. \eqref{eq:_decisionRule1_1} yields 1 when the CI of the parameter $\beta_{s,p_s}$ is fully above the one of $\beta_{s,p_s'}$ and consequently their CIs only allow an isotonic pattern; -1 when it is fully below pointing to an antitonic pattern; and 0 when there exists an overlap, meaning that both monotonicity directions are still possible.

Each result of \eqref{eq:_decisionRule1_1}, denoted as $d_{s,p_s,p_s',\tilde{c}}$, can be understood as an indicator of the relative position of the confidence interval of the parameter $\beta_{s,p_s}$ compared to the one of $\beta_{s,p_s'}$, $\forall p_s'<p_s$ and $p_s \in \{2,3,\ldots,q_s\}$, belonging to the same ordinal predictor $s$ and given a $100\tilde{c}\%$ confidence level. As this is a pairwise comparison, there exist $q_s(q_s-1)/2$ indicators  for each ordinal predictor $s$. Equation \eqref{eq:_decisionRule1_2_V2} uses these indicators to classify the monotonicity direction of an ordinal predictor as a whole at a particular $\tilde{c}$.

The three steps MDC procedure has the following structure: 
\begin{description}
\item [Step 1] 
Set $\tilde{c}$ at a relatively high $100\tilde{c}\%$ confidence level, say $0.99$, $0.95$ or $0.90$, and apply the MDC \eqref{eq:_decisionRule1_2_V2} to assign the subindexes $s$ either to the set $\mathcal{I}$ or $\mathcal{A}$ defined in Section \ref{sec:lkelihood_lagrange}. Therefore, $\mathcal{I}_1=\{s : d_{s,\tilde{c}}=\text{isotonic}\}$ and $\mathcal{A}_1=\{s : d_{s,\tilde{c}}=\text{antitonic}\}$, where $\mathcal{I}_1$ and $\mathcal{A}_1$ denote the isotonic and antitonic sets resulting from the step 1 respectively. In addition, define $\mathcal{B}_1=\{s : d_{s,\tilde{c}}=\text{both}\}$ and $\mathcal{N}_1=\{s : d_{s,\tilde{c}}=\text{none}\}$. If $(\mathcal{I}_1 \cup \mathcal{A}_1)= \mathcal{S}$, then all the ordinal predictors' monotonicity directions have been decided, and there is no need to continue with the MDC procedure. Otherwise, the following step is used for the remaining cases only, $(\mathcal{B}_1 \cup \mathcal{N}_1)$.
\item [Step 2] Consider the set of ordinal predictors $\{s:s\in(\mathcal{B}_1 \cup \mathcal{N}_1)\}$ and apply the MDC \eqref{eq:_decisionRule1_2_V2} in an iterative manner while varying the confidence level $100\tilde{c}\%$. A decrease/increase of $\tilde{c}$ reduces/enlarges the range of the CIs of the parameter $\beta_{s,p_s}$ $\forall s\in (\mathcal{B}_1\cup \mathcal{N}_1)$ and $p_s\in\{2,3,\ldots,q_s\}$. These changes in $\tilde{c}$ produce different effects on the classification depending on whether $s\in \mathcal{B}_1$ or $s\in \mathcal{N}_1$, which must be used as follows:
\begin{enumerate}[label=(\alph*)]
\item For each $s\in \mathcal{B}_1$, the second step is to gradually decrease $\tilde{c}$ while applying the decision rule \eqref{eq:_decisionRule1_2_V2} using a new confidence level $\tilde{c}_s'$ instead of $\tilde{c}$, obtaining $d_{s,\tilde{c}_s'}$. The level of $\tilde{c}_s'$ must be gradually decreased until either a pre-specified minimum confidence level referred to as tolerance level $\tilde{c}_s'^{*}$ is reached, with $0<\tilde{c}_s'^{*}<\tilde{c}$, or the ordinal predictor $s$ is classified as either isotonic or antitonic by $d_{s,\tilde{c}_s'}$.
\item Conversely, for each $s\in \mathcal{N}_1$, gradually increase $\tilde{c}$ while applying the MDC \eqref{eq:_decisionRule1_2_V2} using a new confidence level $\tilde{c}_s''$ obtaining $d_{s,\tilde{c}_s''}$. The level of $\tilde{c}_s''$ must be gradually increased until either a higher confidence level referred to as tolerance level $\tilde{c}_s''^{*}$ is reached, with $\tilde{c}<\tilde{c}_s''^{*}<1$, or the ordinal predictor $s$ is classified as either isotonic or antitonic by $d_{s,\tilde{c}_s''}$.
\end{enumerate}
Finally, $\mathcal{I}_2=\mathcal{I}_1\cup \{s : d_{s,\tilde{c}_s'}=\text{isotonic or }d_{s,\tilde{c}_s''}=\text{isotonic}\}$ and $\mathcal{A}_2=\mathcal{A}_1\cup\{s : d_{s,\tilde{c}_s'}=\text{antitonic}$ or $d_{s,\tilde{c}_s''}=\text{antitonic}\}$, where the subindex of $\mathcal{I}_2$ and $\mathcal{A}_2$ denotes results from the second step. After completing the second step, if $(\mathcal{I}_2 \cup \mathcal{A}_2)=\mathcal{S}$, then it is not necessary to continue with step 3 and the MDC procedure ends. If $(\mathcal{I}_2 \cup \mathcal{A}_2)\subset \mathcal{S}$, then the third and final step must be carried out.
\item [Step 3] Fit $2^{\#\{s:s\notin (\mathcal{I}_2\cup \mathcal{A}_2)\}}$ models accounting for possible combinations of monotonicity directions of the ordinal predictors that were not classified as `isotonic' or `antitonic', i.e., those in the set $\{s:s\notin (\mathcal{I}_2\cup \mathcal{A}_2)\}$, and choose the best model based on some optimality criterion, such as the maximum likelihood as used here.
\end{description}

In general, the MDC procedure describes two levels of decision. The first one is provided by step 1, where a confidence level is applied to all ordinal predictors by the use of a single parameter $\tilde{c}$. The second one is in step 2, where each ordinal predictor $s\in(\mathcal{B}_1 \cup \mathcal{N}_1)$ is classified based on its own confidence level. Step 2 allows to classify predictors that were not classified based on the fixed initial confidence level. 

In step 2, classifying more parameter estimates' patterns with $s\in \mathcal{B}_1$ as either isotonic or antitonic requires a gradual reduction of the confidence level. The tolerance levels $\tilde{c}_s'^{*}$ and $\tilde{c}_s''^{*}$ determine the leeway allowed for the confidence levels in order to enforce a decision. The choice of these may depend on the number of ordinal variables; if the number is small, running step 3 may not be seen as a big computational problem, and it may not be necessary to enforce many decisions in step 2.
The tolerance level $\tilde{c}_s'^{*}$ should not be too low, less than 0.8, say, because it is not desirable to make decisions based on a low probability of occurrence. 

For those $s\in \mathcal{N}_1$ in step 2, the researcher does not face such a trade-off, because greater confidence levels could increase (not decrease) the number of new isotonic or antitonic classifications for those $s\in \mathcal{N}_1$. 

It is important to reduce (or increase) the confidence level in step 2 in a gradual manner, by 0.01 or 0.005, say, for each iteration. If the chosen intervals in the sequence of confidence levels to be assessed are too thick without assessing intermediate levels, then, for an ordinal predictor $s\in \mathcal{B}_1$, it is possible to switch its classification from `both' to `none' instead of updating it from `both' to either `isotonic' or `antitonic'. Conversely, the class of an ordinal predictor $s\in \mathcal{N}_1$ could change from `none' to `both'. The thinner the intervals in the sequence of confidence levels to be assessed are, the less likely to switch from `both' to `none' or `none' to `both' is. However, in some specific cases, there still is a probability of having such an undesired class change.

The researcher may also be interested in exploring other monotonicity directions rather than those resulting from the MDC procedure proposed here, although the maximum likelihood attained by the MDC procedure would not be reached. In this case, the correspondence of each ordinal predictor $s$ to either $\mathcal{I}$ or $\mathcal{A}$ should simply be enforced when constructing $\mathbf{C}$, the matrix of constraints, as described in Section \ref{sec:lkelihood_lagrange}.

In order to illustrate the MDC procedure, we consider a particular example of model \eqref{eq:Model_eq} with four ordinal predictors only ($t=4$ and $v=0$),
where $q_1=3$, $q_2=4$, $q_3=5$, $q_4=6$, and $k=4$, i.e., $j=1,2,3$. The parameters are chosen to be $\alpha_1=-1$, $\alpha_2=-0.5$, and $\alpha_3=-0.1$; and
\begin{align*}
\boldsymbol{\beta}_1'&=(1.0,1.5),\\
\boldsymbol{\beta}_2'&=(0.1,0.2,0.25),\\
\boldsymbol{\beta}_3'&=(-0.02,-0.04,-0.041,-0.05), \text{ and}\\
\boldsymbol{\beta}_4'&=(-0.2,-0.3,-0.31,-0.35,-0.36).
\end{align*}
These parameters represent a situation in which all covariates are monotonic, with the elements of $\boldsymbol{\beta}_1'$ and $\boldsymbol{\beta}_2'$ being isotonic, and those of $\boldsymbol{\beta}_3'$ and $\boldsymbol{\beta}_4'$ antitonic patterns. Given monotonicity, the higher the distances between adjacent parameters are, the clearer the monotonicity direction is. In this illustration, these distances were chosen to make the monotonicity direction clear for the first ordinal predictor only and less clear for the remaining ones, $s=3$ being the most unclear and challenging case because all of its parameters show little distance between adjacent categories and consequently from zero.

\begin{figure*}
	\centering
	\includegraphics[width=1.0\textwidth]{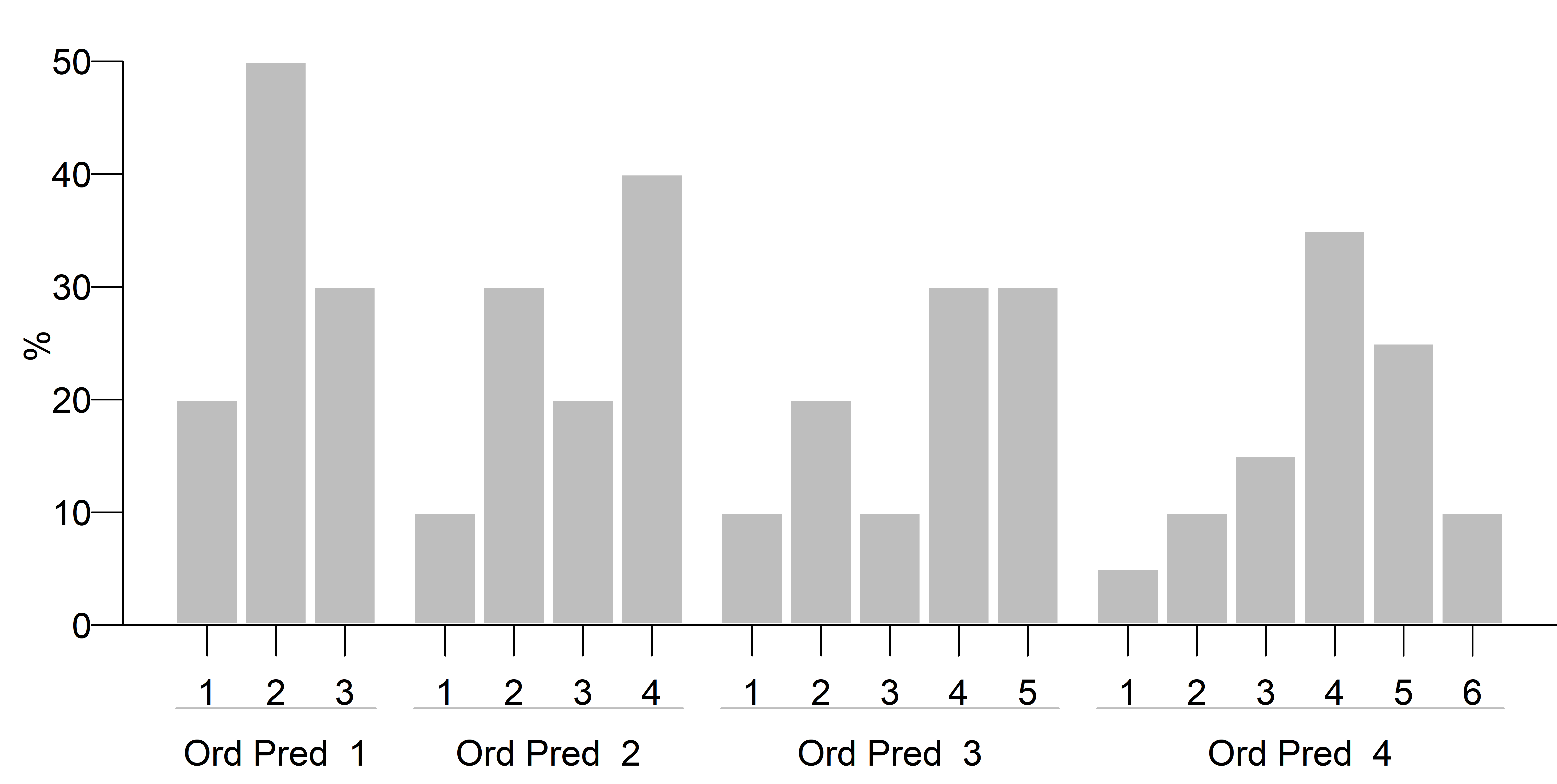}
	\caption[Distributions of simulated ordinal predictors.]{Distributions of simulated ordinal predictors.}
	\label{fig:OrdPredDistributionsExample1}
\end{figure*}

The 2,000 simulated observations of the ordinal predictors were obtained from the population distributions shown in Figure \ref{fig:OrdPredDistributionsExample1}.

Using this simulated data set, an unconstrained version of the model was fitted to obtain the parameter estimates and their standard errors, with which a confidence interval can be computed for any level of $\tilde{\alpha}$ using equation \eqref{eq:CI}.

\begin{figure*}
	\centering
	\includegraphics[width=1.0\textwidth]{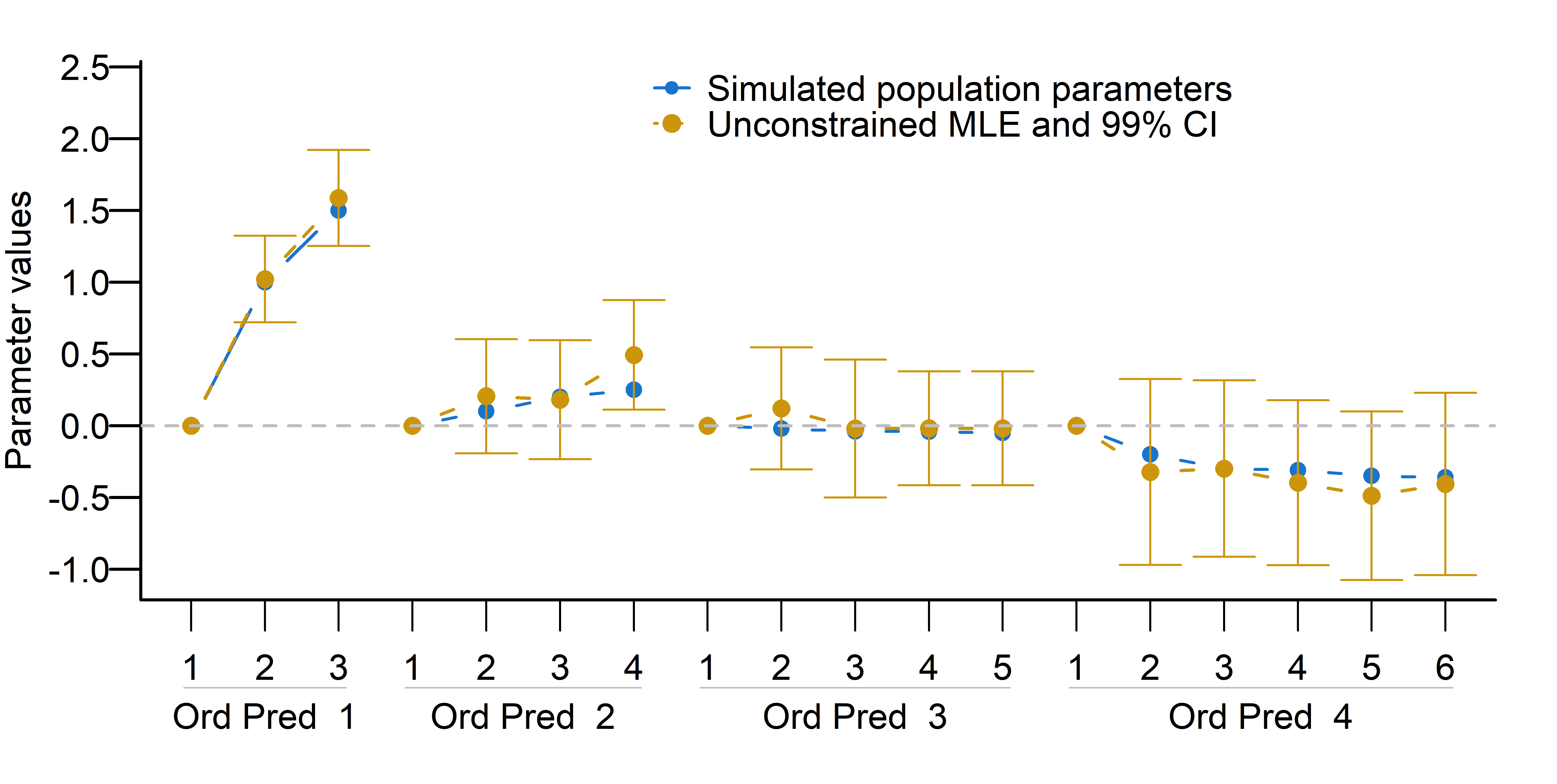}
	\caption[Parameters of ordinal predictors' categories and their unconstrained estimates with 99\% confidence intervals.]{Parameters of ordinal predictors' categories and their unconstrained estimates with 99\% confidence intervals.}
	\label{fig:MonotonicityDirectionsAndCIExample1}
\end{figure*}

For the first step of the MDC procedure, the confidence level was set at a high $\tilde{c}=0.99$. The resulting confidence intervals allowed to classify the first and second OP as `isotonic', $\mathcal{I}_1=\{1,2\}$, and the remaining two patterns of parameter estimates as `both', $\mathcal{B}_1=\{3,4\}$. Figure \ref{fig:MonotonicityDirectionsAndCIExample1} shows that the latter two ordinal predictors allowed both directions of monotonicity, which is the reason why they were not classified as `antitonic'.
The second step was applied over each ordinal predictor $s \in \mathcal{B}_1=\{3,4\}$ using the same tolerance level, $\tilde{c}_3'^{*}=\tilde{c}_4'^{*}=0.8$. For $s=3$, it was not possible to classify its pattern as `antitonic' before reaching the tolerance level. Therefore, it remained as `both'. For $s=4$, the procedure was applied until reaching $\tilde{c}_s'=0.96$, where the fourth OP was classified as `antitonic'. Now, $\mathcal{I}_2=\{1,2\}$ and $\mathcal{A}_2=\{4\}$. As no monotonicity direction was identified for the third OP, two models were fitted in step 3 of the MDC procedure, one treating the third OP as `isotonic' and the other one as `antitonic'. Finally, the model with the highest log-likelihood was selected as the final one.

The procedure successfully classified the ordinal predictors $s=1,2,3,4$ as `isotonic', `isotonic', `antitonic', and `antitonic', respectively, despite the fact that the unconstrained parameter estimates of the last three are not monotonic. Furthermore, it reduced the number of possible models to be fitted from 16 to 2 while making decisions based on individual confidence levels of 96\% or greater.

As shown in Figure \ref{fig:MonotonicityDirectionsAndCIExample1}, it is not easy to classify cases like $s=3$ where all the parameter estimates are close to zero and their confidence intervals are big enough to make the monotonicity direction classification infeasible for any reasonable tolerance level. In this case, the tolerance level would have needed to be set at $\tilde{c}_3'^{*}\le 0.53$ had we wanted the MDC procedure to classify the third ordinal predictor as either `isotonic' or `antitonic'. In fact, when doing so, the MDC makes a mistake and classifies it as `isotonic'. This relationship between low tolerance levels and misclassification is the main reason why the procedure needs to start with a relatively high confidence level $\tilde{c}_s$ and then gradually decrease it until reaching a reasonable tolerance level if necessary.

In cases like $s=3$, one option is to remove this variable from the model because all of the CIs associated to it contain zero even if we choose a tolerance level lower than 0.80, which we consider too low. Removing this variable would have allowed us to fit just one model instead of two. However, removing variables may not be good if the aim is to achieve a model with optimal predictive power.

\section{A monotonicity test} \label{sec:MonoTest}
The MDC procedure assists the decision on the choice of an appropriate monotonicity direction assumption for each OP when fitting model \eqref{eq:Model_eq}, but it is not a formal monotonicity test. It relies on the analysis of multiple pairwise comparisons of confidence intervals with flexibly chosen confidence levels without caring about the simultaneous error probability.

When analysing the monotonicity assumption on the parameters associated to an OP $s$, the Bonferroni correction method can be used to construct a formal monotonicity test for an OP. The Bonferroni correction method allows to compute a set of confidence intervals achieving at least a $100(1 - \alpha_s^*)\%$ confidence level simultaneously (see \cite{miller1981simultaneous}, p. 67, and \cite{bonferroni1936teoria}), which is the probability that all the parameters are captured by the confidence intervals simultaneously. For a given ordinal predictor $s$ and a pre-specified $\alpha_s^*$, if each one of the $q_s-1$ confidence intervals is built with a $100(1-\alpha_s^* /(q_s-1))\%$ confidence level, then the simultaneous confidence level will be at least $100(1-\alpha_s^*)\%$.

The null hypothesis ``$H_0:$ The parameters $\{\beta_{s,p_s}: p_s=1,2,\ldots,q_s\}$ are either isotonic or antitonic'' ($0\leq\beta_{s,2}\leq\beta_{s,3} \cdots \leq \beta_{s,q_s}$ (isotonic) and $0\geq\beta_{s,2}\geq\beta_{s,3} \cdots \geq \beta_{s,q_s}$ (antitonic)) is tested against the alternative ``$H_1:$ The parameters $\{\beta_{s,p_s}: p_s=1,2,\ldots,q_s\}$ are neither fully isotonic nor fully antitonic'' for a given OP $s$, and setting $\beta_{s,1}=0$ as in previous sections. 

For a given ordinal predictor $s$, and taking advantage of the ordinal information provided by its categories, it is then checked whether all the confidence intervals simultaneously are compatible with monotonicity. 

In order to identify whether a pair of confidence intervals of $\beta_{s,p_s}$ is incompatible with monotonicity, a slight modification of equations \eqref{eq:_decisionRule1_2_V2} and \eqref{eq:_decisionRule1_1} is used. Now, instead of the confidence level $\tilde{c}$, those equations use $\tilde{b}=1-\alpha_s^*/(q_s-1)$. Therefore, the monotonicity test for an ordinal predictor $s$ is 
\begin{align}
T_{s,\tilde{b}} = 
     \begin{cases}
       \text{\textit{reject }}H_0  &\quad\text{if }\mathcal{D}_{s,\tilde{b}}\supseteq\{-1,1\} \\
       \text{\textit{not reject }}H_0  &\quad\text{otherwise} \label{eq:decisionRule2_bonferroni}
     \end{cases}
\end{align}
where $\mathcal{D}_{s,\tilde{b}}=\{d_{s,p_s,p_s',\tilde{b}}\}$ is defined as the set of distinct values resulting from using equation \eqref{eq:_decisionRule1_1} for the ordinal predictor $s$ considering each confidence interval with a $100\tilde{b}\%$ confidence level (instead of $100\tilde{c}\%$) in order to achieve a simultaneous confidence level of at least $100(1-\alpha_s^*)\%$ for the parameters associated to the OP $s$.

If $T_{s,\tilde{b}}=\text{\textit{reject }}H_0$, then the parameters associated to the ordinal predictor $s$ are not compatible with the monotonicity assumption with a simultaneous confidence level of at least $100(1-\alpha_s^*)\%$, where $\tilde{b}=1-\alpha_s^*/(q_s-1)$. 

When applying this monotonicity test to the four OPs of the illustration discussed in Section \ref{sec:DecisionRule} and using a pre-specified $\alpha_s^*=0.05$, all the OPs were found to be compatible with the monotonicity assumption.

For a given pre-determined significance level of $\alpha_s^*$ (say 0.1, 0.05 or 0.01), the Bonferroni correction will often be very conservative, and the more conservative, the higher the number of ordinal categories being involved in the monotonicity test. This is because each confidence interval is computed with an individual confidence level of $100(1-\alpha_s^*/(q_s-1))\%$. Therefore, a higher $q_s$ implies larger ranges of the intervals, making the test more likely to not reject $H_0$.

In order to show some results for the monotonicity test with OPs for which their association with the response variable is truly non-monotonic, consider a setting for model \eqref{eq:Model_eq} with two OPs only ($t=2$ and $v=0$), where $q_1=4$, $q_2=5$, and $k=4$, i.e., $j=1,2,3$. The parameters for the intercepts are $\alpha_1=-1$, $\alpha_2=-0.5$, and $\alpha_3=-0.1$; and the true sets of parameters of the OPs 1 and 2 represent non-monotonic associations, being $\boldsymbol{\beta}_1'=(0.4,1.7,0.8)$ and $\boldsymbol{\beta}_2'=(-0.25,-0.7,-0.05,0.40)$. The distributions among categories of OPs 1 and 2 are the same as the ones described in Figure \ref{fig:OrdPredDistributionsExample1} for OPs 2 and 3 correspondingly, and the number of observations is 2,000.

After fitting the new unconstrained model on 1,000 simulated data sets and testing for monotonicity, the null hypothesis was rejected in 84.9\% of the data sets for the OP 1 and in 84.5\%  for the second OP, in both cases with $\alpha^*_s=0.05$.

\section{Dropping constraints and variable selection} \label{sec:Constraints}
\subsection{Dropping monotonicity constraints using the monotonicity test} \label{sec:MonoTestConstraints}
The MDC procedure described in Section \ref{sec:DecisionRule} implies that the parameter estimates of all OPs are restricted to be monotonic. However, the researcher could be interested in imposing monotonicity on a subset of OPs depending on statistical evidence to support this decision, such as the one provided by the monotonicity test.

The monotonicity test could be used as a complementary tool to the MDC procedure in order to assist the estimation process. If the researcher is open to the possibility of not imposing the monotonicity constraints on some OPs, then he/she could first test monotonicity on each one of them and then perform the MDC procedure on all those ordinal predictors where the null hypothesis of the monotonicity test was not rejected. Under this scenario, in case that monotonicity is rejected for an OP, it would be more prudent to fit unconstrained estimates on the parameters associated to it. Therefore, such an OP should not be part of $\mathcal{S}$, the set of OPs to be constrained, but rather part of the non-ordinal predictors, considering it at the nominal scale level.

\subsection{Dropping monotonicity constraints using the MDC procedure } \label{sec:MDCConstraints}
When dropping the monotonicity constraint for some of the OPs is considered as a feasible option, then not only the approach introduced in Section \ref{sec:MonoTestConstraints} could be used, but also an alternative one is proposed in this section. As in the previous section, consider the case where the researcher might also want to explore whether the monotonicity assumption holds for all of the OPs or for a subset of them, but now using a less conservative approach than the monotonicity test. 

An adjusted version of the MDC procedure described in Section \ref{sec:DecisionRule} allows to drop the monotonicity assumption for some OPs. There are only two adjustments, one in step 2.b and the other one in step 3. The first one is to set $\tilde{c}_s''^{*}=\tilde{c}$, i.e., the tolerance level for each OP $s\in \mathcal{N}_1$ is set to be the same as the confidence level chosen in step 1. Therefore, the second step is not performed on any ordinal predictor $s\in \mathcal{N}_1$. The second modification is to apply step 3 over the possible combinations of monotonicity directions of the ordinal predictors that were classified as `both' by the end of step 2, i.e., the number of models to be fitted is now $2^{\#\{s : d_{s,\tilde{c}_s'^{*}}=\text{both}\}}$ instead of $2^{\#\{s:s\notin (\mathcal{I}_2\cup \mathcal{A}_2)\}}$. This implies that $\mathcal{S}$, the set of OPs to be constrained, must be updated excluding each ordinal predictor $s\in \mathcal{N}_1$ from the set of monotonicity constraints. Finally, the model should be fitted considering these OPs as nominal scaled variables.

These adjustments are equivalent to consider the first step of the MDC procedure as a filter of OPs to be constrained, where those that are classified as `none' by the end of this step are removed from $\mathcal{S}$ and excluded from steps 2 and 3.

\subsection{Using the MDC procedure for variable selection} \label{sec:MDCSelection}
The parameter estimates' patterns classified as `both' at the end of the second step of the MDC procedure are also of interest. `Both' refers to an ordinal predictor for which all of the parameters associated to its categories have CIs containing zero. Therefore, if this is true even for the CIs evaluated at the tolerance level, an option is to remove such an ordinal predictor from the model of interest and apply the MDC procedure again using the new model. If more than one OP is classified as `both' and there is appetite to drop such variables, then it is advisable to do it in a stepwise fashion such as backward elimination, while checking the results of the MDC procedure in each step, because dropping an OP could affect the monotonicity direction classification of another OP. We will not investigate this in detail here, assuming that the data is rich enough so that variable selection is not required.

Each of the options described in Sections \ref{sec:MDCConstraints} and \ref{sec:MDCSelection}, i.e., dropping monotonicity constraints for those ordinal predictors $s\in \mathcal{N}_1$ and dropping ordinal predictors $\{s : d_{s,\tilde{c}_s'^{*}}=\text{both}\}$, reduces the number of models to be fitted in step 3. If these options are applied to all of the ordinal predictors classified as `both' or `none', then step 3 would be avoided.

\section{Simulations} \label{sec:Simulations}
The model \eqref{eq:Model_eq} with two ordinal and two interval scale predictors,
\begin{align}
\text{logit}[P(y_{i}\leq j | \mathbf{x}_i)]&=\alpha_j+\sum_{p_1=2}^{4}\beta_{1,p_1}x_{i,1,p_1} \nonumber \\
&+\sum_{p_2=2}^{6}\beta_{2,p_2}x_{i,2,p_2}+\beta_{1}x_{i,1}+\beta_{2}x_{i,2},\label{eq:Model_eq_simulations}
\end{align}
where $k=5$, i.e., $j=1,2,3,4$, was fitted for 1,000 data sets simulated as described in Section \ref{sec:DecisionRule} using the following parameters: for the intercepts $\alpha_1=-1.4$, $\alpha_2=-0.4$, $\alpha_3=0.3$, and $\alpha_4=1.1$; for the ordinal predictor's categories $\boldsymbol{\beta}_1'=(0.3,1.0,1.005)$, and $\boldsymbol{\beta}_2'=(-0.2,-1.5,-1.55,-2.4,-2.41)$; and for the interval scale predictors $\beta_1=-0.15$ and $\beta_2=0.25$. The parameters vectors $\boldsymbol{\beta}_1$ and $\boldsymbol{\beta}_2$ are chosen to represent isotonic and antitonic patterns respectively.

The number of simulated observations for each data set is 1,000. The ordinal predictors were drawn from the population distributions used in Section \ref{sec:DecisionRule} for those covariates with the same number of ordinal categories, 4 and 6. The interval scale covariates $x_1$ and $x_2$ were randomly generated from normal distributions, $N(0,1)$ and $N(5,4)$ correspondingly. 

For each one of the 1,000 data sets, model \eqref{eq:Model_eq_simulations} was fitted following four different approaches:
\begin{enumerate}
\item UMLE: Using unconstrained MLE, which implies not imposing monotonicity constraints on any of the predictors.
\item CMLE: Using constrained MLE, which implies imposing monotonicity constraints on all of the OPs as described in Section \ref{sec:DecisionRule}. In this simulation, the MDC procedure was preformed choosing a 90\% confidence level in step 1 ($\tilde{c}=0.90$).
\item CMLE Bonferroni: Using constrained MLE imposing monotonicity constraints on those ordinal predictors to which the null hypothesis of the monotonicity test was not rejected as described in Section \ref{sec:MonoTestConstraints}. In the current setting, the chosen simultaneous significance level for the monotonicity test was $\alpha_s^*=0.05$, for $s=1,2$, and the confidence level for the first step of the MDC procedure was 90\% ($\tilde{c}=0.90$).
\item CMLE filtered: Using constrained MLE imposing monotonicity constraints on those ordinal predictors not classified as `none' by the first step of the MDC procedure as described in Section \ref{sec:MDCConstraints}, with the same pre-specified confidence level of 90\% ($\tilde{c}=0.90$) as in the previous approaches.
\end{enumerate}

After fitting the UMLE, the MDC procedure was performed. Its first step classified the OP 1 as `isotonic' in 100\% of the cases and the OP 2 as `antitonic' in 98.5\%. The remaining 1.5\% correspond to the cases where OP 2 was classified as `none'. However, by the end of the second step, with tolerance level $\tilde{c}_2''^{*}=0.999$, the MDC procedure correctly classified the monotonicity directions for 100\% of the data sets with no need of performing its third step.

When applying the monotonicity test to the OPs of model \eqref{eq:Model_eq_simulations}, the null hypothesis of monotonicity was not rejected in 100\% of the data sets for both OPs with $\alpha_s^*=0.05$.

The few cases representing the 1.5\% of the data sets where the OP 2 was classified as `none' by the first step of the MDC procedure produced some differences between the approaches CMLE and `CMLE filtered'. In contrast, the monotonicity test results did not produce any difference between the approaches CMLE and `CMLE Bonferroni'.

\begin{figure*}
	\centering
	\includegraphics[width=1\textwidth]{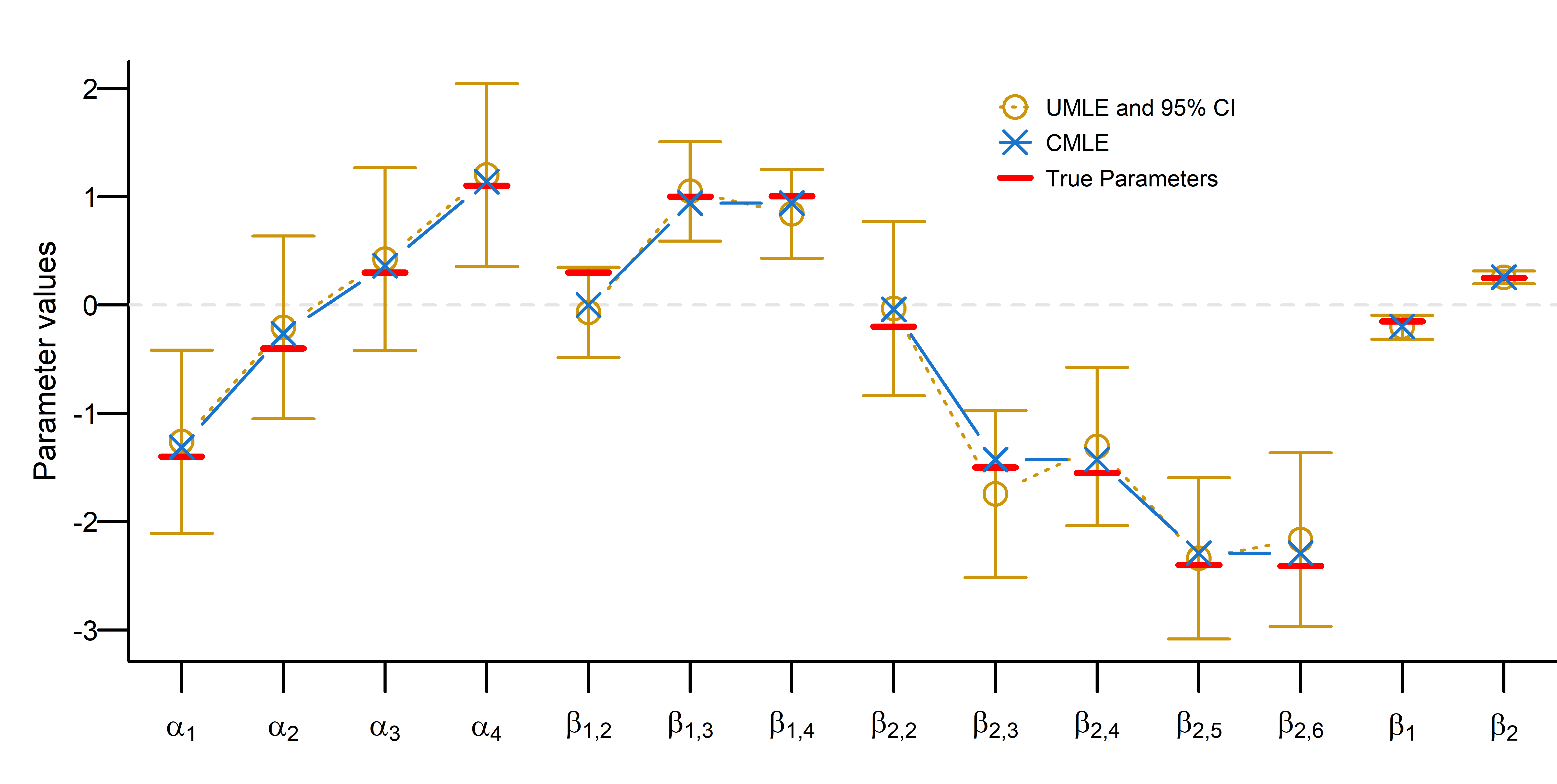}
	\caption[Unconstrained MLE and constrained MLE: An example from simulations.]{An example of unconstrained MLE and constrained MLE from simulations.}
	\label{fig:UMLEvsCMLE_ParEst_Simul_SingularExample}
\end{figure*}

Consider one of the 1,000 data sets for illustration. As shown in Figure \ref{fig:UMLEvsCMLE_ParEst_Simul_SingularExample}, some unconstrained parameter estimates are incompatible with the monotonicity assumptions of this simulated model. Despite the fact that the first ordinal predictor is assumed to be isotonic, the UMLE yields $\hat{\beta}_{1,2}<0$ and $\hat{\beta}_{1,3}>\hat{\beta}_{1,4}$. Similar violations occur with the second ordinal predictor, which is assumed to be antitonic but $\hat{\beta}_{2,3}<\hat{\beta}_{2,4}$ and $\hat{\beta}_{2,5}<\hat{\beta}_{2,6}$. However, the results of the CMLEs impose the monotonicity assumptions, with the estimate for $\beta_{1,2}>0$, the estimate for $\beta_{1,4}$ being slightly greater than the one for $\beta_{1,3}$, and where the estimates for $\beta_{2,4}$ and $\beta_{2,6}$ are slightly lesser than those for $\beta_{2,3}$ and $\beta_{2,5}$ respectively.

The results of approaches `CMLE Bonferroni' and `CMLE filtered' were omitted in Figure \ref{fig:UMLEvsCMLE_ParEst_Simul_SingularExample} because they are exactly the same as the ones of the approach `CMLE' for this particular example. This is because the results of this simulated data set show that the first step of the MDC procedure did not classify OPs 1 or 2 as `none', and the monotonicity test did not reject the null hypothesis of monotonicity for any of these two OPs.

The CMLEs for the parameters that are not related to the ordinal predictors are, in general, not the same as the UMLEs. In this particular example, the parameter estimates associated to both intercepts and interval scale covariates are hardly affected by the monotonicity assumption when comparing the CMLE to the UMLE.

\begin{figure*}
	\centering
	\includegraphics[width=1\textwidth]{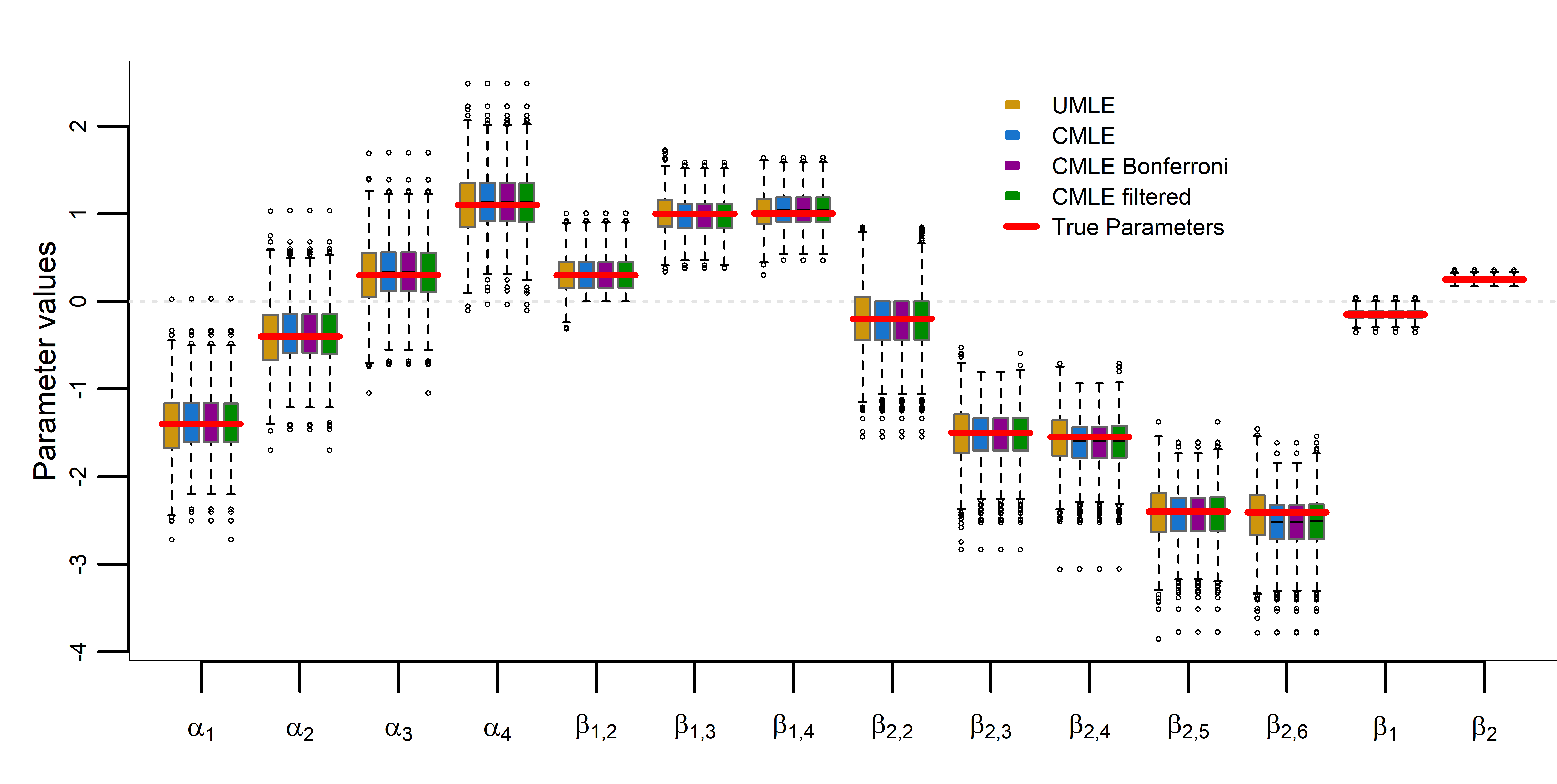}
	\caption[Unconstrained MLE and constrained MLE: Results from simulations for a model with 2 OPs.]{Unconstrained MLE, three methods based on constrained MLE and parameters used for 1,000 simulated data sets with 2 OPs.}
	\label{fig:UMLEvsCMLE_ParEst_Boxpl_Simul}
\end{figure*}

Figure \ref{fig:UMLEvsCMLE_ParEst_Boxpl_Simul} uses boxplots to visualise the distributions of the parameter estimates of the four approaches listed above together with the parameters used in the data generation process for the 1,000 simulations.

The effect of the monotonicity constraints is depicted by the range of values of the approach CMLE for each one of the ordinal predictor categories' parameters, which differ from the UMLEs in two aspects. The first one is that the parameter estimates are forced to take values being compatible with their monotonicity direction, i.e., positive for the isotonic case and negative for the antitonic one. This is why the boxplots of the CMLE for $\beta_{1,2}$ and $\beta_{2,2}$ seem to be truncated at zero as opposed to the UMLE. The second difference is a generalization of the first one. The lower extremes of the CMLE boxplots make their corresponding whiskers shorter than the ones of the UMLE when there is an isotonic relationship, and the same effect occurs for the upper whiskers when the relationship is antitonic. However, this does not produce big differences between the UMLE and CMLE in terms of their median. Regarding the comparison between the approach `CMLE Bonferroni' and the unconstrained MLE, exactly the same conclusions hold. This is because the monotonicity tests did not provide statistical evidence against monotonicity for any OP.

The results of the approach `CMLE filtered' are slightly different from the other constrained approaches because there are 15 cases where the OP 2 was considered as non-monotonic. Therefore, the monotonicity constraints were not imposed on OP 2 when fitting their corresponding models. This is the reason why there are some positive values for the estimate of $\beta_{2,2}$ in the approach `CMLE filtered'.

\begin{figure*}
	\centering
	\includegraphics[width=1\textwidth]{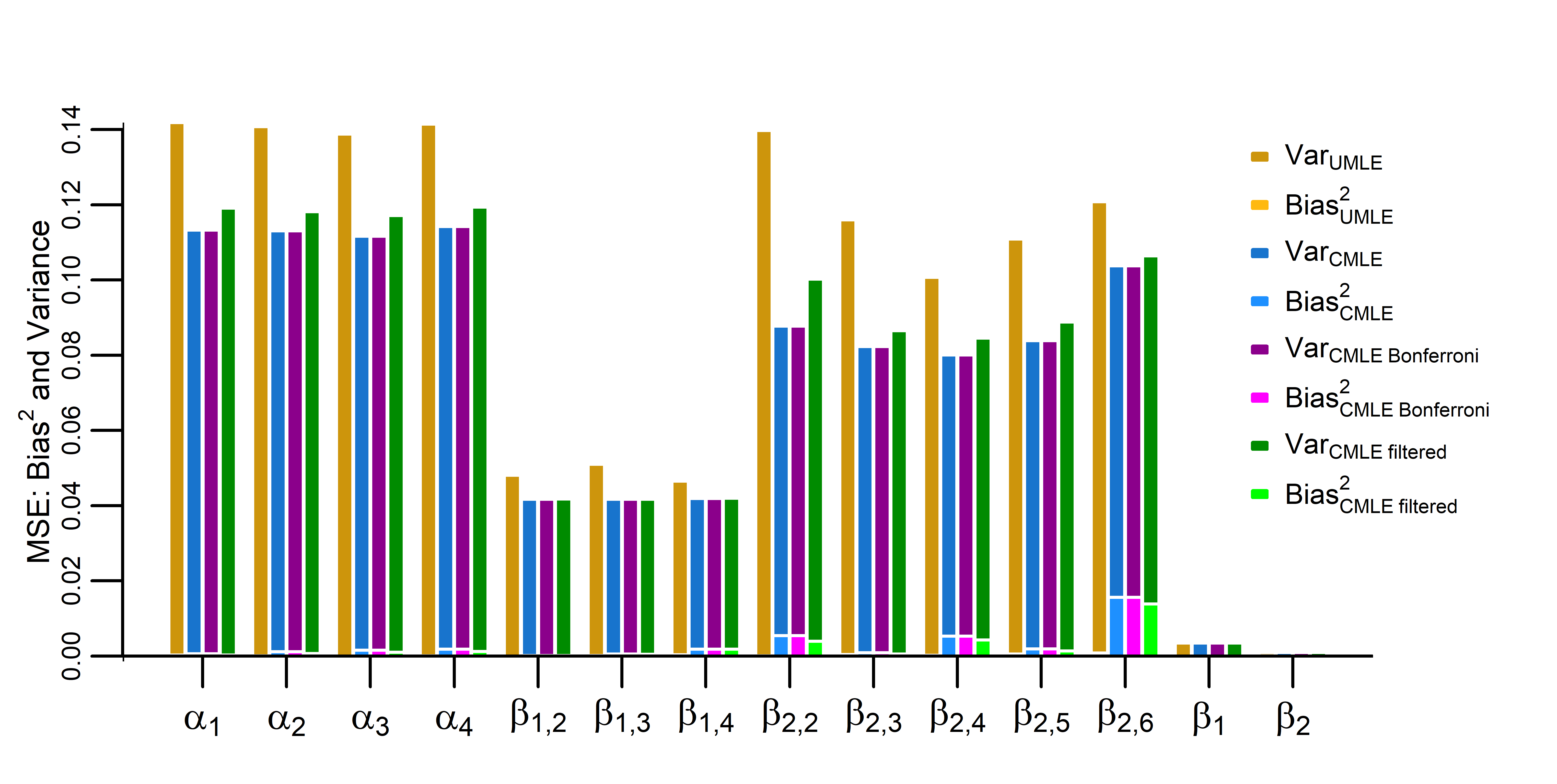}
	\caption[Mean square error for unconstrained and constrained MLEs and its decomposition.]{Mean square error for unconstrained and constrained MLEs and its decomposition.}
	\label{fig:MSE_Simulations}
\end{figure*}

Consider the mean square error (MSE) and its squared bias-variance decomposition to compare the performance of the three different approaches of the CMLE with respect to the one of the UMLE. As shown in Figure \ref{fig:MSE_Simulations}, the total MSE is notably smaller for the CMLE. On average, the CMLE for the intercepts shows a 19.7\% smaller MSE compared to the MSE of UMLE, 13.8\% smaller for the first ordinal predictor set of CMLEs, and 25.1\% smaller for the second one. The same results are for `CMLE Bonferroni' compared to UMLE, and the corresponding figures for `CMLE filtered' versus UMLE are 15.8\%, 13.8\% and 20.3\%.

Despite the fact that the squared bias makes a markedly small contribution to the total MSE (lighter colours in Figure \ref{fig:MSE_Simulations}), it is clearly higher for some constrained parameter estimates, specially for those associated to the second ordinal predictor. The CMLE associated to the sixth category of the second ordinal predictor produced the highest squared bias, which represents 15.0\% of its total MSE (the same for `CMLE Bonferroni' and 13.0\% for `CMLE filtered'). The squared bias of the remaining CMLEs associated to the second OP together with those related to the first OP and the intercepts represent, on average, 2.4\% of the MSE, ranging from 0.2\% to 6.5\%. The corresponding figures for `CMLE filtered' are 1.8\%, 0.2\%, and 5.0\%. Consequently, the MSEs are composed mainly of variances, which are considerably lower not only for the parameters associated to the ordinal predictor categories, but also for the intercepts. In general, the total MSE for the CMLE is always significantly smaller because the squared bias is more than compensated by the much smaller variance.

In the previous simulation, no non-monotonic ordinal predictor was included and its results showed that any approach with CMLE performed better than the unconstrained one. In order to analyse their performance in presence of non-monotonic OPs, consider another simulation of model \eqref{eq:Model_eq}. This time we use an ordinal response with four categories, i.e., $k=4$ and $j=1,2,3$; four ordinal predictors ($t=4$) with $q_1=3$, $q_2=4$, $q_3=5$, and $q_4=6$ categories correspondingly; one interval scale predictor ($v=1$); and $n=500$. The chosen parameters for the intercepts are $\alpha_1=-1.4$, $\alpha_2=-0.1$, and $\alpha_3=1.7$; for OP 1 are $\boldsymbol{\beta}_1'=(0.5,1)$; for OP 2 are $\boldsymbol{\beta}_2'=(-0.65,-0.70,-1.60)$; for OP 3 are $\boldsymbol{\beta}_3'=(0,0,0,0)$; for OP 4 are $\boldsymbol{\beta}_4'=(-0.8,-1.6,-0.6,0.6,1.6)$; and for the interval scale predictor $\beta_1=0.3$. The parameters of the OPs 1 to 4 were chosen to be isotonic, antitonic, zero, and non-monotonic correspondingly. For OP 3, all the parameters were set to zero, and therefore the monotonicity test is expected to not reject monotonicity and the third step of the MDC procedure is expected to increase the number of models to be fitted because its second step should classify it as `both'. Hence, OPs 3 and 4 contribute to challenge the MDC procedure and the monotonicity test respectively.

This model was fitted for 1,000 simulated data sets. The ordinal predictors were drawn from the population distributions showed in Figure  \ref{fig:OrdPredDistributionsExample1}. The interval scale predictor was randomly generated from a normal distribution $N(1,4)$.

In general, the MDC procedure was executed with a $90\%$ confidence level in the first step ($\tilde{c}=0.90$) and tolerance levels $\tilde{c}_s'^{*}=0.85$ and $\tilde{c}_s''^{*}=0.999$ for $s=1,2,3,4$ in the second step. The monotonicity direction results for the three different approaches to the constrained estimation process are the following:
\begin{itemize}
\item CMLE: This approach demands to impose monotonicity restrictions on all of the OPs. Therefore, their monotonicity directions were always established. OPs 1 and 2 were finally classified as `isotonic' and `antitonic' in 100.0\% of the data sets. The OP 3 remained classified as `both' in 64.0\% of the data sets at the end of the second step. Therefore, the third step was completed and OP 3 was classified as `isotonic' in 48.8\% of the data sets and as `antitonic' in the remaining 51.2\%. Finally, the first step classified the OP 4 as `none' in 96.8\% of the data sets, and by the end of the whole procedure it was classified as `isotonic' in 84.1\% of the cases and as `antitonic' in the remaining 15.9\%.
\item CMLE Bonferroni: This approach does not impose monotonicity constraints on those OPs that are non-monotonic according to the monotonicity test results. The null hypothesis of monotonicity was not rejected for OPs 1, 2 and 3 for all of the data sets, with $\alpha^*_1=\alpha^*_2=\alpha^*_3=0.05$. Regarding OP 4, the test did reject $H_0$ in 83.9\% of the data sets with $\alpha^*_4=0.05$. Therefore, the MDC procedure was fully applied to determine the monotonicity direction of OPs 1 to 4 for 16.1\% of the data sets, whereas it was not used to establish the monotonicity direction of OP 4 in the corresponding 83.9\% as it was assumed to be non-monotonic. The final monotonicity direction classifications for the 1,000 data sets were 100\% `isotonic' for OPs 1; 100\%  `antitonic' for OP 2; 48.2\% `isotonic' for OP 3 (51.8\% `antitonic'); and 15.1\% `isotonic', 1.0\% `antitonic' and 83.9\% `none' for OP 4.
\item CMLE filtered: Like in the previous approach, this one imposes monotonicity constraints on those OPs that are assumed as monotonic only. According to the first step of the MDC procedure, the OPs 1 and 2 were compatible with monotonicity for 100\% of the data sets. However, this step classified OPs 3 and 4 as `none' in 0.2\% and 96.8\% of the data sets respectively. Thus, the final monotonicity direction classifications were 100\% `isotonic' for OPs 1; 100\% `antitonic' for OP 2; 47.7\% `isotonic', 52.1\% `antitonic' and 0.2\% `none' for OP 3; and 3.2\% `isotonic' and 96.8\% `none' for OP 4.
\end{itemize}

\begin{figure*}
	\centering
	\includegraphics[width=1\textwidth]{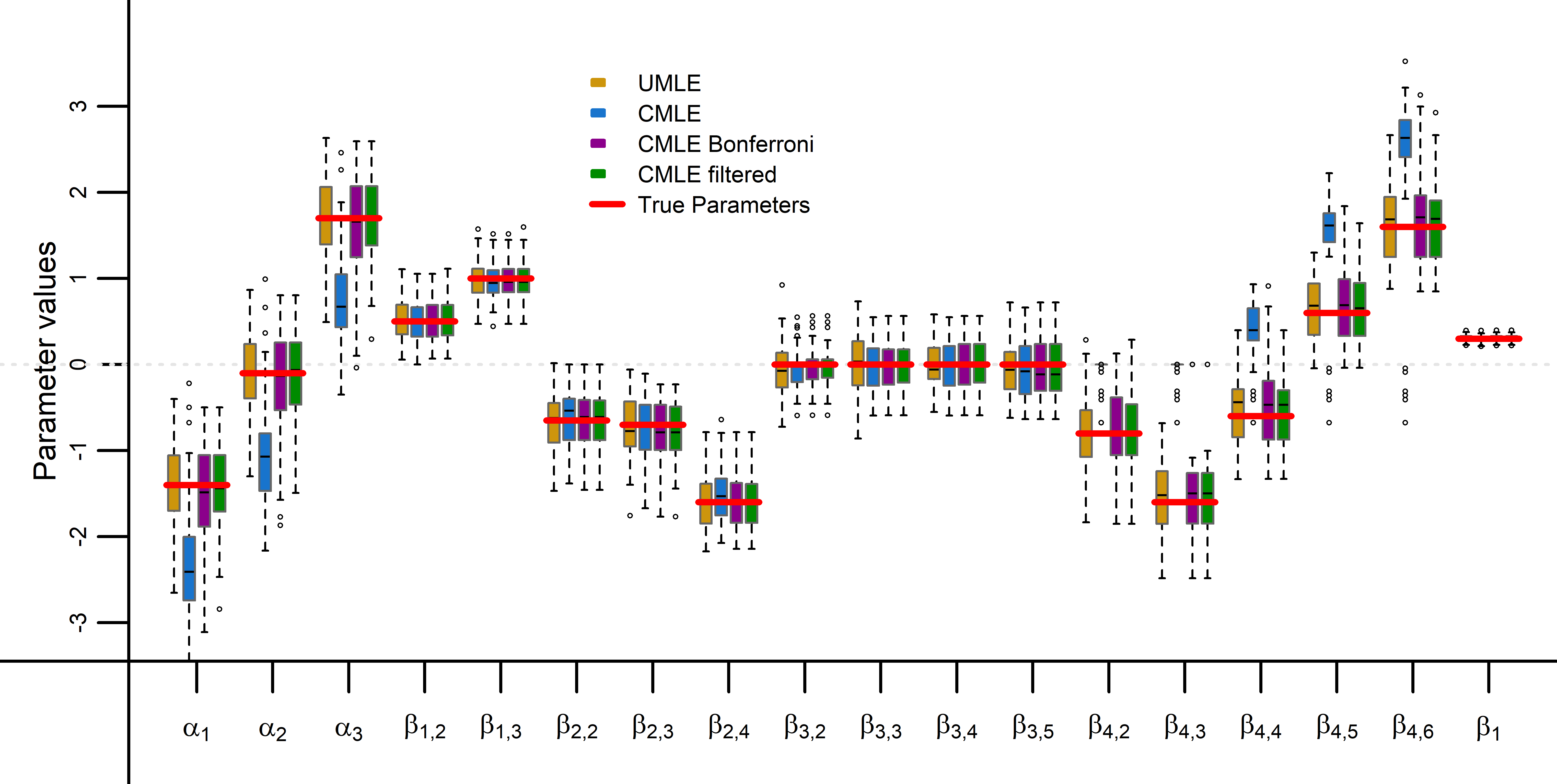}
	\caption[Unconstrained MLE and constrained MLE: Results from simulations for a model with 4 OPs.]{Unconstrained MLE, three methods based on constrained MLE and parameters used for 1,000 simulated data sets with 4 OPs.}
	\label{fig:UMLEvsCMLE_ParEst_Boxpl_Simul2}
\end{figure*}

Figure \ref{fig:UMLEvsCMLE_ParEst_Boxpl_Simul2} shows the results of the unconstrained MLE and the three different approaches for the constrained MLE.

If the researcher is not open to the possibility of dropping monotonicity constraints and there is a non-monotonic ordinal predictor, such as OP 4 in this simulation, then a higher bias is the price of imposing monotonic effects in order to get, for example, easier interpretability for all of the OPs using the approach `CMLE'. This was expected because, for all of the data sets, this approach forced the parameters of OP 4 to be monotonic but this is not its true pattern. This is why the CMLEs for OP 4 (blue boxplots in Figure \ref{fig:UMLEvsCMLE_ParEst_Boxpl_Simul2}) depart from the other approaches and the true parameters, which also produces a high bias in the parameters associated to the intercepts.

Assuming that the researcher is willing to drop monotonicity constraints depending on statistical evidence and given that there was a non-monotonic ordinal predictor, the approaches `CMLE Bonferroni' and `CMLE filtered' provided reasonable results. Their performances in terms of MSE depend on the degree of conservativeness when establishing the set of OPs with non-monotonic effects. The more conservative, the higher MSE compared to the one of UMLE. In this case, the approach `CMLE filtered' resulted to be the best option within the constrained ones. On average, it is the only approach that produced a lower MSE (1.5\% lower) compared to the one of UMLE. The MSE of `CMLE Bonferroni' was, on average, 33.1\% higher than the one of unconstrained MLE, mainly because of OP 4, which was 98.7\% higher. This is because the approach `CMLE Bonferroni' favoured monotonicity of OP 4 for 16.1\% of the data sets, whereas `CMLE filtered' did this for 3.2\% of the data sets only.

\section{Application to quality of life assessment in Chile} \label{sec:RealData}
As an illustration of the the proposed methodology, we analyse the association between a quality of life self-assessment variable (10-point Likert scale) and ordinal and other predictors. The data source is a Chilean survey, the National Socio-Economic Characterisation 2013 (CASEN, from its name in Spanish). This survey retrieves information with the aim of characterising the population of people and households. Our analysis is based on 7,374 householders corresponding to those who live in the capital and reported the quality of life self-assessment.

The survey provides information on several variables, from which the final set of covariates was chosen on the basis of previous research in the field (see for example \cite{di2003macroeconomics, cheung2014assessing, boes2010effect}). The data set analysed in the current section was published by the Ministry of Social Development of Chile and it is available online at: \url{http://observatorio.ministeriodesarrollosocial.gob.cl/casen-multidimensional/casen/basedatos.php}. In order to reproduce the data set used in this section, the steps involved in the data preprocessing are described in Appendix A.

The response variable is a self assessment of the quality of life (QoL). Each respondent was asked to answer the question `Considering everything, how satisfied are you with your life at this moment?'. The possible alternatives were: `1 Completely Unsatisfied', `2',$\ldots$, `9', `10 Completely Satisfied'. 

The model was fitted with ordinal, ratio and nominal scale covariates. For the ordinal and nominal scale ones, the first category to be mentioned is considered as the baseline. The ordinal covariates are \textit{Educational Level (Edu)} with categories `Not Educated', `Primary', `Secondary', and `Higher'; \textit{Income Quintile (Inc)} with levels from `Q1' to `Q5' where `Q5' represents the highest income; \textit{Health Status (Hea)}, a health self-assessment reported as ordinal Likert scale from 1 to 7, with 7 being the best possible status; \textit{Overcrowding (Ove)}, which is an index representing the number of people living in the household per bedroom, with categories `Not Overcrowded' for less than 2.5, `[2.5,3.5)', `[3.5,5.0)', and `5.0 or more'; and \textit{Children (Chi)}, a grouped version of the number of people under 15 years old living in the household, with categories `0',`1', `2', `3', and `4 or more'. The ratio scale variable is \textit{Age}. The nominal scale ones are \textit{Activity (Act)}, with categories `Economically Inactive', `Unemployed', and `Employed'; and \textit{Sex} with categories `Male' and `Female'. Therefore, the set of ordinal predictors is $\mathcal{S}=\{Edu, Inc,Hea,Ove,Chi\}$.

Each set of parameter estimates associated to the ordinal predictors in $\mathcal{S}$ was classified as either `antitonic' or `isotonic'. The interpretation for the relationship between an ordinal predictor and the response variable with `antitonic' pattern is that the further away an ordinal category is from its baseline, the smaller $P(y_i\leq j|\mathbf{x}_i)$ is, i.e., the probability of self-assessing \textit{QoL} in the $j$th category or smaller. In other words, `antitonic' patterns mean that higher categories of ordinal variables are associated to more probability of self-assessing \textit{QoL} in a higher part of the scale. The inverse interpretation applies for `isotonic' patterns.

\begin{figure*}
	\centering
	\includegraphics[width=1\textwidth]{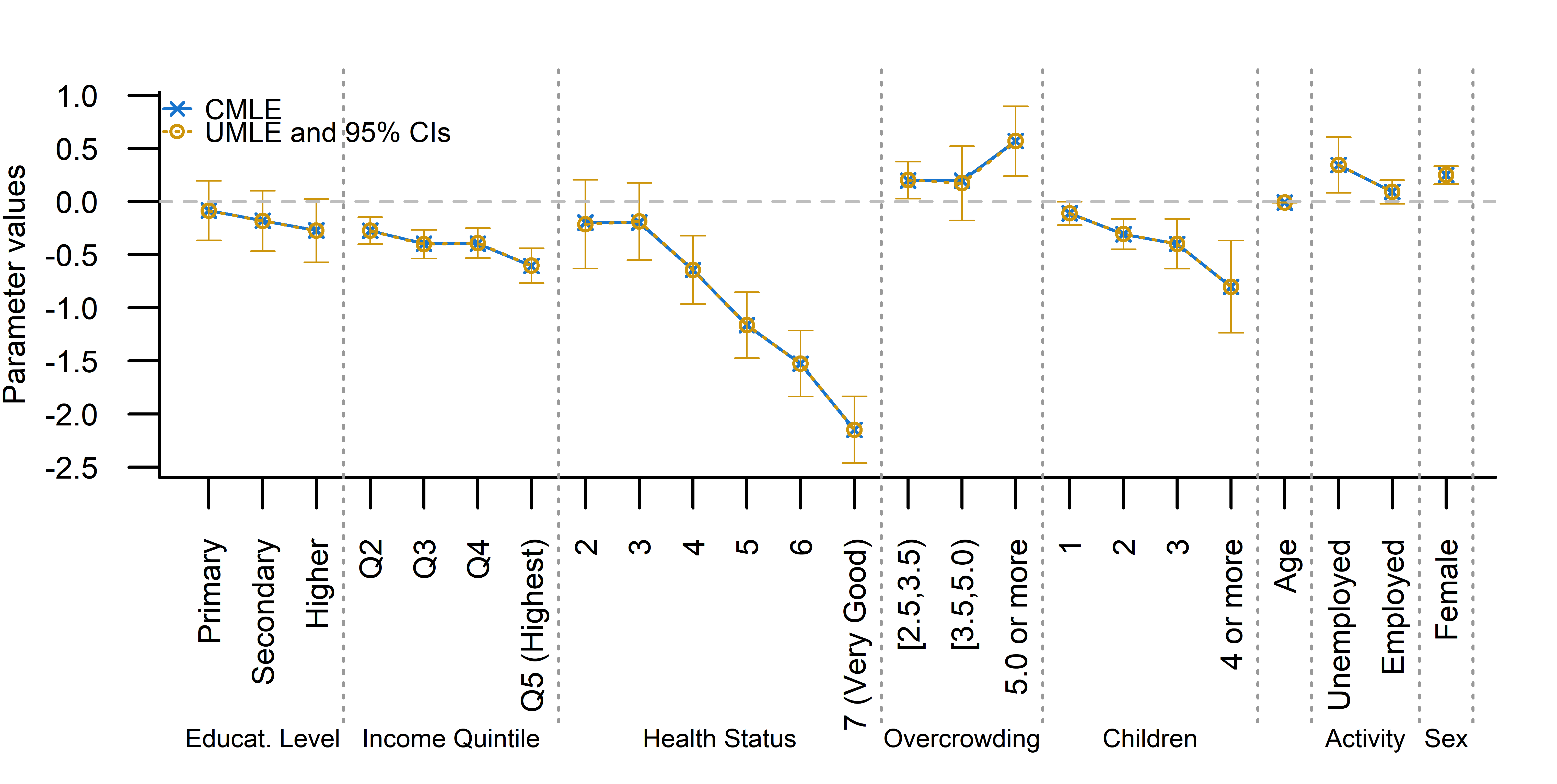}
	\caption[An application of the constrained regression model in real data (intercepts omitted).]{CMLEs and UMLEs for a model applied on real data with an ordinal response, ordinal predictors and others. Intercepts parameter estimates omitted. The 95\% confidence intervals correspond to the UMLEs.}
	\label{fig:RealDataSetExample2}
\end{figure*}

An unconstrained version of the model \eqref{eq:Model_eq}, the proportional odds cumulative logit model, was fitted to obtain the parameter estimates and their standard errors. The unconstrained parameter estimates and their 95\% confidence intervals are shown in Figure \ref{fig:RealDataSetExample2}. These results seem to be consistent with the monotonicity assumption for all the OPs. Therefore, the assumption of monotonicity was imposed on all of them and the `CMLE' approach was used.

The monotonicity directions were established by using the MDC procedure. With a 95\% individual confidence level ($\tilde{c}=0.95$), the MDC procedure classified the sets of parameters associated to three ordinal variables as `antitonic' in its first step (\textit{Income Quintile}, \textit{Health Status}, and \textit{Children}), whereas \textit{Overcrowding} was classified as `isotonic' and \textit{Educational Level} as `both'. The latter decision was made because all of its 95\% individual CIs contain zero as shown in Figure \ref{fig:RealDataSetExample2}. There was no ordinal predictor classified as `none' by the end of the first step. Therefore, there was no need of making a decision on whether dropping the monotonicity constraints for those classified as `none' or not. Hence, $\mathcal{A}_1=\{Inc,Hea,Chi\}$, $\mathcal{I}_1=\{Ove\}$, and $\mathcal{B}_1=\{Edu\}$.

\textit{Educational Level} was the only variable in the MDC procedure's second step. To perform this step, a tolerance level of 0.9 was set together with steps of 1\% when gradually decreasing the confidence level starting from the one analysed in step one, 95\%. As a result, the \textit{Educational Level} ordinal variable was classified as `antitonic' at the 92\% confidence level for each confidence interval.

There was no need to execute the third step of the MDC procedure because all of the monotonicity directions were established earlier. All the ordinal predictors were finally classified as `antitonic' except for \textit{Overcrowding}, which was classified as `isotonic'. Therefore, only one model was fitted.

We also used the monotonicity test described in Section \ref{sec:MonoTest} as a complementary assessment of the monotonicity assumptions. Its results are consistent with the MDC procedure in the sense that it did not reject the null hypothesis of monotonicity for any of the ordinal predictors with $\alpha^*=0.05$.

Some of the parameter estimators resulting from the unconstrained estimation are not in line with the monotonicity assumption. The third UMLE of \textit{Income Quintile} and the second one of \textit{Health Status} are greater than their previous adjacent one, which is not compatible with their monotonicity direction because they both are assumed to be `antitonic'. In the opposite direction, the same happens with the second UMLE of \textit{Overcrowding}. For instance, keeping all the other variables constant, an improvement in the \textit{Income Quintile} from `Q3' to `Q4', i.e., an increment in the income level, increases the probability of self-assessing \textit{QoL} in lower categories of the scale, according to the UMLE. The same happens with \textit{Health Status}, for which changes from `2' to `3', i.e., improving the health status, seemingly increases the probability of reporting a low self-assessment of \textit{QoL}. Despite the fact that the true parameters are unknown, these particular unconstrained results are counterintuitive. Therefore, it is reasonable to think that these may have been the result of random variation, and to impose the monotonicity assumption here if we want to avoid misinterpretations.

For the OP \textit{Educational Level}, the UMLE allows both positive and negative values in all confidence intervals, but after having classified this OP as antitonic, with the baseline parameter fixed at zero and using the CMLE, all further parameters can only be negative.

In general, the UMLEs are compatible with a monotonic association between ordinal predictors and the response variable. However, the parameter estimates produce some violations of monotonicity. The CMLEs avoid these, and allow for a simpler and more consistent interpretation.

\section{Conclusions} \label{sec:Conclusions}
We propose a constrained regression model for an ordinal response with ordinal predictors,which can involve other types of predictors. The information provided by the category ordering of the ordinal predictors is used appropriately for ordinal data, rather than ignoring it (assuming categories as nominal) or overstating it as interval.

Each set of parameters associated to an ordinal predictor's categories can be enforced to be monotonic in our procedure, which decides automatically whether associations are isotonic or antitonic. The monotonicity direction classification procedure can classify variables not only as isotonic or antitonic, but also as compatible with both monotonicity directions or none, and the researcher may sometimes prefer to leave out variables compatible with both direction and zero parameters, and to drop the monotonicity constraint for variables incompatible with either direction, which can easily be done within the framework presented here.

The MDC relies on the choice of a pre-specified range of confidence levels between $\tilde{c}_s'^{*}$ and $\tilde{c}_s''^{*}$, but the regression model itself does not require a tuning parameter and does deliver monotonic parameter estimates, unlike the penalised version in \cite{tutz2014rating}, which pushes parameters in the direction of monotonicity but does not necessarily achieve it.

A monotonicity test is proposed to assess the validity of the monotonicity assumption for every ordinal predictor. This checks whether the set of confidence intervals belonging to the parameters of an ordinal predictor is compatible with monotonicity or not. As this is based on the Bonferroni correction of confidence levels, it can be expected to be very conservative, and more powerful tests can probably be developed. This is left to future work.

Three different approaches for the estimation method are proposed depending on whether the researcher wants to impose monotonicity constraints on all of the OPs or some subset of them. For the first case, the MDC procedure is fully applied, and for the second case, the two remaining approaches differ in the way they identify the subset of OPs on which the monotonicity assumption is not imposed, one uses the monotonicity test, and the other one executes a modified version of the MDC procedure.

The CMLE for the real data application proved to be a sensible solution because it enabled a consistent interpretation for the ordinal variables' categories, which would not have been the case for the UMLE. 

Asymptotic theory for the CMLE is a matter of ongoing research. This would enable us to make inference about the parameters in the constrained model. 

We intend to produce one single \texttt{R} package containing the implementation of the constrained regression model for ordinal predictors discussed in Section \ref{sec:Model}; the MDC procedure with access to the results from each one of its steps (Section \ref{sec:DecisionRule}); and the monotonicity test discussed in Section \ref{sec:MonoTest}.

\bibliographystyle{apalike}

\newpage
\begin{appendices}
\section{Real data reproducibility}

In Section \ref{sec:RealData} of the paper, we present a real data application to quality of life self assessment in Chile. In order to assist reproducibility, we provide the criteria used in the data preprocessing stage to get the final data set from the raw data that is publicly available online at: \url{http://observatorio.ministeriodesarrollosocial.gob.cl/casen-multidimensional/casen/basedatos.php}.

\subsection{Response variable and sample definition}
The response variable is a self assessment of the
quality of life (QoL), the name of this variable in the original data set is \textbf{\texttt{r20}} and its possible values are integers from 1 to 10, representing the possible answers: `1 Completely Unsatisfied', `2',\ldots , `9', `10 Completely Satisfied' correspondingly.

The sample is defined as those householders who live in the capital and reported the quality of life self assessment. In the original data set, householders are identified with the value 1 of the variable \textbf{\texttt{pco}}, whereas the capital corresponds to  \textbf{\texttt{region=13}} and the valid responses of QoL lie between 1 and 10.

\subsection{Predictors}

\subsubsection{Ordinal predictors}
\begin{description}
\item [Educational Level:] This variable takes into account the educational level, years of schooling, and whether the householder knows how to read and/or write. All these variables are treated in the following sequential steps:
\begin{enumerate}
\item Variable \textbf{\texttt{educ}} is grouped into four categories: values 0 and 99 in ``Not educated'', values 1 and 2 in ``Primary'', values from 3 to 6 in ``Secondary'', and from 7 to 12 in ``Higher''.
\item Those classified as ``Secondary'' are moved to ``Primary'' if their years of schooling are less than 9 (variable \textbf{\texttt{ESC}}$>$9).
\item Those classified as ``Not educated'' and with \textbf{\texttt{educ}}=99 are moved to ``Primary'' if their years of schooling are more than 0 (variable \textbf{\texttt{ESC}}$>$0).
\item Those classified as ``Primary'' and with \textbf{\texttt{educ}}=99 are moved to ``Secondary'' if their years of schooling are more than 8 (variable \textbf{\texttt{ESC}}$>$8).
\item Those classified as ``Secondary'' and with \textbf{\texttt{educ}}=99 are moved to ``Higher'' if their years of schooling are more than 12 (variable \textbf{\texttt{ESC}}$>$8).
\item Those classified as ``Secondary'' or ``Higher'', and declared that they do not know how to read and/or write (variable \textbf{\texttt{e1}} is 2, 3, or 4), are moved to ``Primary''.
\item Those classified as ``Not educated'' and with \textbf{\texttt{educ}}=99 (value for `do not know/do not answer') are removed from the sample (28 cases, 0.37\%).
\end{enumerate}
\item [Income Quintile:] Raw variable \textbf{\texttt{QAUTR\_MN}} is used.
\item [Health Status:] Variable \textbf{\texttt{s16}} is used. Values from 1 to 7 are considered only and those observations with value 99 (value for `do not know') are removed from the sample (36 cases, 0.48\%).
\item [Overcrowding:] Variable \textbf{\texttt{hacinamiento}} is used. Values from 1 to 4 are considered only and those observations with value 9 (value for `NA') are removed from the sample (21 cases, 0.28\%).
\item [Children:] This is a special case, we use the whole data set and a dummy variable to identify those people under 15 years old. Then grouping by the house identifier called \textbf{\texttt{folio}} we get the number of children by house. We incorporate this information back in the sample of householders living in the capital and reported the quality of life self assessment. Finally, we grouped the number of children when it is greater than or equal to 4.
\end{description}

\subsubsection{Non-Ordinal predictors}
None of the non-ordinal predictors was transformed. The name of variables ``Age'', ``Activity'', and ``Gender'' are \textbf{\texttt{edad}}, \textbf{\texttt{activ}}, and \textbf{\texttt{sexo}}, correspondingly.
\end{appendices}

\end{document}